\documentclass[preprint,superscriptaddress,showkeys,showpacs,
tightenlines,fleqn,nofootinbib,prd,byrevtex]{revtex4} 
\usepackage{graphicx}
\usepackage{bm}
\usepackage{epstopdf}
\begin{document}      
\preprint{YITP-07-66}
\preprint{Inha-NTG-01/2008}
\title{Pion electromagnetic charge radii and $\rho$-meson mass shift\\
 at finite density}       
%-------------------------------------------------
\author{Seung-il Nam}
\email[E-mail: ]{sinam@yukawa.kyoto-u.ac.jp}
\affiliation{Yukawa Institute for Theoretical Physics (YITP), Kyoto
University, Kyoto 606-8502, Japan} 
%-------------------------------------------------
\author{Hyun-Chul Kim}
\email[E-mail: ]{hchkim@pusan.ac.kr}
\affiliation{Department of Physics, Inha University, Incheon 402-751,
Republic of Korea} 
\affiliation{Department of Physics and Nuclear Physics \& Radiation
  Technology Institute (NuRI), 
Pusan National University, Busan 609-735, Republic of Korea} 
%---------------------------------------------------------------------
\date{January 2008}
\begin{abstract}  
We investigate the pion electromagnetic charge radii and mass dropping
of the $\rho$ meson at finite density.  We first calculate the pion
charge radii within the framework of the nonlocal chiral quark model
from the instanton vacuum both at zero and finite densities.  In order
to relate the change of the pion charge radius to that of the
$\rho$-meson mass at finite density, we employ the vector-meson
dominance for the pion electromagnetic form factor.  It turns out that
the pion charge radius is getting larger as the quark chemical
potential increases.  As a result, the $\rho$-meson mass falls off as
the quark chemical potential grows and is found to be dropped by about
$10\%$ at normal nuclear matter density: $m^*_{\rho}/m_{\rho}\approx 1 -
0.12\,\mu_B/\mu_0$, which is compatible to the results of the
measurement at the KEK recently. 
\end{abstract}
%---------------------------------------------------------------------
\maketitle
%---------------------------------------------------------------------
\section{Introduction}
%---------------------------------------------------------------------
It has been widely known that properties of hadrons undergo
changes at finite temperature and/or density.  In particular, the
modification of vector mesons in medium plays an essential role in
understanding the properties of the QCD vacuum under hot and/or dense 
environment (see, for example, the following references and those
therein~\cite{Cassing:1999es,Rapp:1999ej,Brown:2002is}), since there 
may be a possible connection between their in-medium modification and
the restoration of chiral symmetry of which spontaneous breakdown
provides a mechanism for the mass generation of most hadrons.  In
particular, it is of great interest to see the in-medium change of the
$\rho$ meson, since it is strongly coupled to the $\pi\pi$ isovector
channel and may be observed in dilepton productions
($\pi^+\pi^-\to\rho\to l^+l^-$) in heavy-ion collision because of the
fact that the dileptons produced from the $\rho$ meson can convey
information on the mass shift of the $\rho$ meson without any
contamination due to the strong interaction.  

In fact, there are already experimental results for changes of the 
$\rho$-meson properties in medium: The CERES/NA45 collaboration has
investigated low-mass electron pair productions in various collisions 
~\cite{Agakishiev:1995xb,Agakishiev:1997au,Adamova:2002kf}.  The NA60
collaboration has recently announced that the averaged $\rho$-meson
spectral function shows a strong broadening but no shift in
mass~\cite{Arnaldi:2006jq}.  The CLAS collaboration has reported that
the medium effects on the $\rho$-meson mass are not
observed~\cite{Djalali:2007zz,CLAS:2007mga} and some widening of the
$\rho$-meson decay width is seen for the Fe-Ti target.  However, the
E325 experiment at the KEK 12-GeV Proton Synchrotron has measured the
invariant mass spectra of $e^+e^-$ pairs in $p+A$ reactions and has
found that the $\rho$-meson mass is dropped by about
$9\,\%$~\cite{Naruki:2005kd}.  The broadening of the $\rho$-meson
width does not seem to fit the E325 data.  The 
STAR collaboration has also reported a possible shift of the
$\rho$-meson peak in the $\pi^+\pi^-$ channel in Au+Au and $p$+$p$
collisions at $\sqrt{s}=200\,\mathrm{GeV}/c^2$~\cite{Adams:2003cc}.
The TAGX collaboration has observed an in-medium modification of the
$\rho$-meson spectral function in $^2\mathrm{H}$, $^3\mathrm{He}$, and
$^{12}\mathrm{C}(\gamma,\pi^+\pi^-)$ reactions~\cite{Huber:2003pu}.  

The in-medium change of the $\rho$ meson has been extensively studied in
many different theoretical approaches.  Brown and
Rho~\cite{Brown:1991kk} first proposed a possible dropping of the   
$\rho$-meson mass, introducing the effective scalar glueball field in
order to make the effective chiral Lagrangian consistent with the QCD
scaling property.  Ref.~\cite{Brown:1991kk} estimated the $\rho$-meson
mass in medium dropped by about $20\,\%$.  Subsequently, Hatsuda 
and Lee have shown its mass shifted by about
$18\,\%$~\cite{Hatsuda:1991ez}, based on the QCD sum rules.  Carter et  
al.~\cite{Carter:1995zi}, on the contrary, have found that it turns
out much smaller, i.e. around $3\,\%$, since the glueball field is
rather stable in medium in their effective Lagrangian approach.
Klingl et al.~\cite{Klingl:1997kf}, however, have investigated the
in-medium $\rho$-meson properties in the QCD sum rules but have
concluded that the change of the $\rho$ meson in medium cannot be
interpreted simply by the mass dropping of the $\rho$ meson alone.
Refs.~\cite{Rapp:1995zy,Post:2000qi} have shown in the study of the
$\rho$-meson spectral function that the width broadening is more
important to interpret the corresponding experimental data.  Thus, no 
consensus has been reached, either experimentally or theoretically, as
yet on how the $\rho$ meson experiences the change in medium.

In the present work, we want to investigate the electromagnetic charge
radii of the pion at finte density but $T=0$ within the framework of
the nonlocal chiral quark model (N$\chi$QM) from the instanton vacuum, 
based on which the mass shift of the $\rho$ meson will be evaluated,
associating with the vector meson dominance (VMD) in a
phenomenological approach.  We already studied the electromagnetic
(EM) form factor of the pion within the same
framework~\cite{Nam:2007gf}.  Since the instanton vacuum at finite 
density was investigated in Ref.~\cite{Carter:1998ji} some years ago,
we can immediately use the formalism of Ref.~\cite{Carter:1998ji} to
extend our previous work into finite density.  Thus, we first 
consider in this work the pion EM form factor at finite quark chemical
potentials ($\mu_q$) and its charge radii.  Having shown how much the pion
charge radius is modified in medium, we try to translate its change at 
finite density into the corresponding shift of the
$\rho$-meson mass in medium with the help of the VMD.  

The present work is organized as follows: In Section II, we make a
brief introduction of the N$\chi$QM and its application to the nonzero 
$\mu_q$. The parameterized pion EM form factor in terms of the VMD is 
introduced in order to determine the mass dropping of the $\rho$
meson.  The numerical results with discussions are given in
Section III.  The final Section is devoted to the summary and conclusion.

%---------------------------------------------------------------------%
\section{Pion Electromagnetic form factor}
%---------------------------------------------------------------------%
We start with the definition of the pion EM form factor in the
space-like region:
\begin{equation}
 \label{eq:emff}
\langle \pi^+(p_f)|j^{\rm EM}_{\mu}(0) |\pi^+ (p_i)\rangle
=(p_f + p_i)_{\mu} F_{\pi} (Q^2), 
\end{equation}
where $|\pi^+\rangle$ stands for the positively charged pion state
($\pi^+ (u\bar{d})$).  The $p_i$ and $p_f$ represent the initial 
and final on-shell momenta for the pion, satisfying
$p^2_i=p^2_f=m^2_{\pi}$.  The pion mass $m_{\pi}$ is taken to be 140
MeV for numerical input.  The momentum transfer is defined as
$q^2=(p_f-p_i)^2$ and $Q^2=-q^2>0$ in the space-like region.  Here,
the local EM current in flavor SU(2) is defined in Euclidean space as
follows:     
\begin{equation}
\label{eq:emcurrent}
j^{\rm EM}_{\mu}(x) = iq^\dagger(x) \hat{Q} q (x) = 
i\frac{2}{3}u^{\dagger}(x)\gamma_{\mu}u(x)-i\frac{1}{3}  
d^{\dagger}(x)\gamma_{\mu}d(x)
\end{equation}
with the charge operator $\hat{Q}=\mathrm{diag}(2/3,-1/3)$.  
Note that all calculations are performed in Euclidean space,
since we are interested in the form factor in the space-like region.
The EM form factor satisfies the normalization condition at $Q^2=0$ by 
charge conservation: $F_{\pi}(0)=1$.  In Ref.~\cite{Nam:2007gf}, we
have already computed the pion EM form factor in free space, employing
the nonlocal chiral quark model (N$\chi$QM) from the instanton
vacuum~\cite{Diakonov:1983hh,Diakonov:1985eg,Diakonov:2002fq}.  A
great virtue of the N$\chi$QM from the instanton vacuum lies in the
fact that there are only two parameters, namely, the average instanton 
size $\bar \rho\approx \frac{1}{3}\, 
\mathrm{fm}$ and average inter-instanton distance $\bar R\approx
1\, \mathrm{fm}$.  The normalization scale of this approach can be
defined by the average size of instantons and is approximately
equal to $\bar{\rho}^{-1}\approx 600$  MeV. The values of
the $\bar \rho$ and $\bar R$ were estimated phenomenologically in
Ref.~\cite{Shuryak:1981ff} as well as theoretically in 
Ref.~\cite{Diakonov:1983hh,Diakonov:2002fq,Schafer:1996wv}.  
Furthermore, it was confirmed by various lattice simulations of
the QCD vacuum \cite{Chu:vi,Negele:1998ev,DeGrand:2001tm}. Moreover, 
lattice calculations of the quark propagator~\cite{Faccioli:2003qz,
Bowman:2004xi} are in a remarkable agreement with 
that of Ref.~\cite{Diakonov:1983hh}.  A recent lattice simulation
with the interacting instanton liquid model obtains $\bar
\rho\approx 0.32\,\mathrm{fm}$ and $\bar R\approx 0.76
\,\mathrm{fm}$ with the finite current quark mass $m$ taken into
account~\cite{Cristoforetti:2006ar}.  The N$\chi$QM turns out to be
very successful in describing properties of hadrons in the low-energy
regime.  

However, the presence of the nonlocal interaction between quarks and
pseudo-Goldstone bosons breaks the gauge invariance for the
external vector fields.  Since the pion EM form factor involves the
vector current, we need to deal with this problem.  While 
Ref.~\cite{Pobylitsa:1989uq} proposed a systematic way as to how the
conservation of the N\"other current is restored, one has to handle
the integral equation.  Refs.~\cite{Musakhanov:2002xa,Kim:2004hd}
derived the light-quark partition function in the presence 
of the external gauge fields.  With this gauged partition function, it
was shown that the low-energy theorem for the transition from
two-photon state to the vacuum via the axial anomaly was
satisfied~\cite{Musakhanov:2002xa}.  Moreover, the gauged effective
chiral action has been shown to describe well a great deal of
mesonic properties such as the meson distribution
amplitudes and kaon semileptonic decay form factors~\cite{Nam:2006sx,
Nam:2007fx}.  As shown in Refs.~\cite{Musakhanov:2002xa,Kim:2004hd},
if the external vector fields are weak enough, the gauged effective
chiral action in the chiral limit can be written in terms of the
covariant derivative $D$:  
\begin{eqnarray} 
\label{eq:ECA} 
\mathcal{S}_{\mathrm{eff}} = -\mathrm{Sp} \ln \left[i\rlap{/}{D} +
  i\sqrt{M_q(iD)} U^{\gamma_5}\sqrt{M_q(iD)}\right], 
\end{eqnarray} 
where the functional trace ${\rm Sp}$ runs over space-time, color,
flavor and Dirac spaces.  The $iD$ is the covariant derivative
expressed as $i\partial+e_q V$, where $V$ stands for the external EM
gauge field.  The $U^{\gamma_5}$ denotes the nonlinear Goldstone boson
fields defined as:   
\begin{equation}
U^{\gamma _{5}}= \exp\left(i\gamma_5 \tau\cdot\pi/f_\pi\right),
\end{equation} 
where $f_{\pi}$ stands for the pion decay constant: $f_\pi=93$
MeV.  Note that the pion EM form factor derived from
Eq.~(\ref{eq:ECA}) can be shown to satisfy obviously the
Ward-Takahashi identity.  We refer to Ref.~\cite{Nam:2007gf} for the
detailed formalism of how to derive the pion EM form factor from
Eq.~(\ref{eq:ECA}).  

Now, we are in a position to introduce the quark chemical potential
($\mu_q$) in the model in order to calculate the pion EM form factor
in medium.  Since we consider the case of the chiral limit, isospin
symmetry is automatically taken into account:
$m_{\mathrm{u}}=m_{\mathrm{d}}=0$.  Thus, we need to consider only the
isoscalar quark chemical potential.  In the present work, we
closely follow Ref.~\cite{Carter:1998ji}.  Moreover, we take into
account the gauge invariance in the presence of the external EM field, 
as mentioned before.  The relevant effective chiral action at finite
density can be written as follows: 
\begin{eqnarray} 
\label{eq:ECA1} 
{\cal S}_{\rm eff}[\mu_q]=-{\rm Sp}\ln\left[i\rlap{/}{D}-
i\rlap{/}{\mu} + i\sqrt{{\cal M}(iD)} U^{\gamma_5}\sqrt{{\cal M}(iD)}
\right], 
\end{eqnarray} 
%EQUATION<<<
where $\mu=(0,0,0,\mu_q)$ denotes the quark chemical potential as a
four vector.  The momentum-dependent quark mass, ${\cal M}(iD)$,
modified by the nonzero quark chemical potential, can be obtained by
the Fourier transform of the following modified Dirac equation for the
quark zero mode in the instanton vacuum:
\begin{eqnarray} 
\label{eq:MDE} 
\left[i\rlap{/}{\partial}-i\rlap{/}{\mu}+\hat{A}_{I\bar{I}}
\right]\Psi^{(0)}_{I\bar{I}} = 0.
\end{eqnarray} 
%EQUATION<<<
The momentum-dependent quark mass at finite density has been
obtained already in Ref.~\cite{Carter:1998ji}.  However, it is
rather complicated to use that derived in Ref.~\cite{Carter:1998ji},
so that in the present work we make a parameterization for
$\mathcal{M}$ in the following simple form:
\begin{eqnarray}
\label{eq:MFD}
\mathcal{M}(i\partial,\mu)&=& \mathcal{M}_0
\left[\frac{2\Lambda^2}{(i\partial-i\mu)(i\partial-i\mu)+2\Lambda^2}
\right]^2={\cal M}_0{\cal F}^2(i\partial,\mu),
\end{eqnarray}
where ${\cal F}$ denotes the quark form factor which is related to 
Eq.~(\ref{eq:MDE}).  Note that there is a caveat in using
Eq.~(\ref{eq:MFD}): Such a simple parameterization is only valid at
dilute density ($\mu_q\lesssim 300$ MeV)~\cite{Nametal}.  Otherwise,
the results would become unstable.  

In Fig.~1, we compare the parameterized quark form factor
given in Eq.~(\ref{eq:MFD}) (solid curves) with that from 
Ref.~\cite{Carter:1998ji} (dashed curve) at $\mu_q=100$ MeV and   
$|{\bm p}|=200$ MeV.
\begin{figure}[t]
\includegraphics[width=10cm]{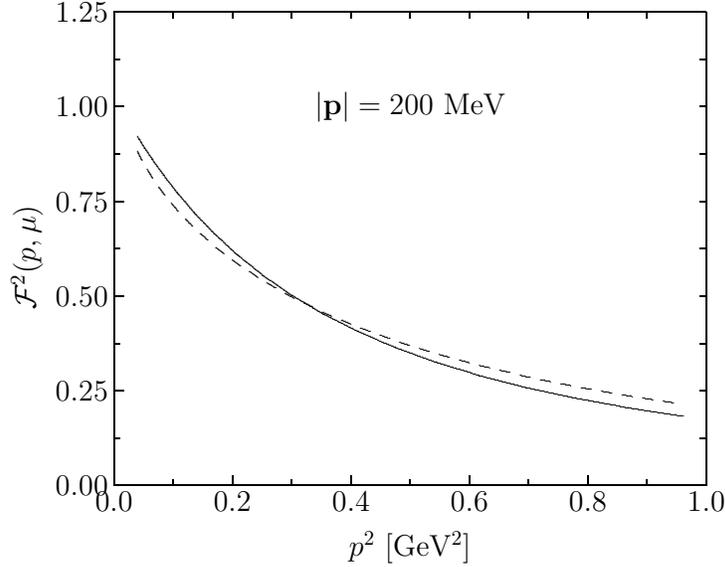}
\label{fig:1}
\caption{Comparison of ${\cal F}^2(p,\mu)$ in Eq.~(\ref{eq:MFD}) with 
that of Ref.~\cite{Carter:1998ji}.  The solid curve depicts the
parameterized form factor, while the dashed one draws that of
Ref.~\cite{Carter:1998ji}.  The form factors are calculated at
$\mu_q=100$ MeV and $|{\bm p}|=200$ MeV.} 
\end{figure}
As shown in Fig.~1, the simple parameterized form factor
reproduces that of Ref.~\cite{Carter:1998ji} qualitatively well.  The 
cut-off mass $\Lambda$, which is taken to be the inverse of the
avergage instanton size, can be consistently determined by the
self-consistency equation:   
\begin{eqnarray}
\label{eq:NV}
\frac{N}{V}=4N_c\int^{\infty}_{-\infty}\frac{d^4p}{(2\pi)^4}
\frac{\mathcal{M}^2(p,\mu)}{(p+i\mu)^2+\mathcal{M}^2(p,\mu)}. 
\end{eqnarray}
%EQUATION<<<
Although there must be corrections due to the presence of the quark
chemical potential, they are very small as far as we stay below the
region below normal nuclear matter density ($\mu_q\lesssim300$
MeV).  Hence, we take the vacuum value $N/V\approx(200\,{\rm
  MeV})^4$ and it will give $\mathcal{M}_0\approx350$ MeV.  Following the
same formalism as in Ref.~\cite{Nam:2007gf} and introducing the
nonzero quark chemical potential, we obtain the following local and
nonlocal contributions to the pion EM form factor in
Eq.~(\ref{eq:emff}) at $\mu_q\ne0$:  
%EQUATION>>>
\begin{eqnarray}
\label{eq:local}
&&F^{*\, \mathrm{local}}_{\pi}=\sum_{\rm flavor}
\frac{8e_qN_c}{(p_i\cdot q+2m^2_{\pi})}
\int^{\infty}_{-\infty}\frac{d^4k}{(2\pi)^4}
\Bigg[\frac{\sqrt{{\cal M}_b{\cal M}_c}({\cal M}_ck_{bd}+{\cal
    M}_bk_{cd})} {2(k^2_b+{\cal M}^2_b)(k^2_c+{\cal M}^2_c)}
\nonumber\\
&&\hspace{1cm}+\frac{{\cal M}_a\sqrt{{\cal M}_b{\cal M}_c}
(k_{ab}k_{cd}+k_{ac}k_{bd}
-k_{bc}k_{ad}+{\cal M}_a{\cal M}_ck_{bd}+{\cal M}_a{\cal M}_bk_{cd} 
-{\cal M}_c{\cal M}_ck_{ad})}{(k^2_a+{\cal M}^2_a)(k^2_b+{\cal M}^2_b)
(k^2_c+{\cal M}^2_c)}\Bigg],
\nonumber\\
\label{eq:Nonlocal}
&&F^{*\, \mathrm{nonlocal}}_{\pi}
=\sum_{\rm flavor}
\frac{8e_qN_c}{(2p_i\cdot q+M^2_{\pi})}
\int^{\infty}_{-\infty}\frac{d^4k}{(2\pi)^4}\Bigg[
\frac{\sqrt{{\cal M}_b{\cal M}_c}(\sqrt{{\cal M}_c}\hat{\cal 
    M}_{bd}-\sqrt{M_b}\hat{\cal M}_{cd}) 
(k_{bc}-{\cal M}_b{\cal M}_c)}{(k^2_b+{\cal M}^2_b)(k^2_c+{\cal
  M}^2_c)} 
\nonumber\\
&&\hspace{1cm}+\frac{{\cal M}_a\sqrt{{\cal M}_b{\cal M}_c}(\sqrt{{\cal 
 M}_b} \hat{\cal M}_{cd}-\sqrt{{\cal M}_c}\hat{\cal M}_{bd})({\cal
M}_ck_{ab} + {\cal M}_bk_{ac}-{\cal M}_ak_{bc}+{\cal M}_a{\cal
M}_b{\cal M}_c)} {(k^2_a+{\cal M}^2_a)(k^2_b+{\cal M}^2_b)(k^2_c+{\cal
M}^2_c)} 
\nonumber\\
&&\hspace{1cm}+\frac{\sqrt{{\cal M}_a{\cal M}_c}
\left[\sqrt{{\cal M}_b}\hat{\cal M}_{ad}-\sqrt{{\cal M}_a}\hat{\cal
    M}_{bd}\right] (k_{ac}+{\cal M}_a{\cal M}_c)}
{2(k^2_a+{\cal M}^2_a)(k^2_c+{\cal M}^2_c)}
\nonumber\\
&&\hspace{1cm}+\frac{\sqrt{{\cal M}_a{\cal M}_b}
\left[\sqrt{{\cal M}_c}\hat{\cal M}_{ad}-\sqrt{{\cal M}_a}\hat{\cal
    M}_{cd}\right] (k_{ab}+{\cal M}_a{\cal M}_b)}
{2(k^2_a+{\cal M}^2_a)(k^2_b+{\cal M}^2_b)}\Bigg],
\end{eqnarray}
%EQUATION<<<
where ${\cal M}_{\alpha}={\cal M}(k_{\alpha})$. The relevant momenta
are defined as $k_a=k+\mu-p_i/2-q/2$, $k_b=k+\mu+p_i/2-q/2$,
$k_c=k+\mu+p_i/2+q/2$ and $k_d=p_i$. Note
that all the momenta depend on $\mu_q$.  We use the following
abbreviations:    
$k_{\alpha\beta}=k_{\alpha}\cdot k_{\beta}$ and $\hat{\cal
  M}_{\alpha\beta}$ is defined as the  derivative of the  
dynamical quark mass: 
%EQUATION>>>
\begin{equation}
\hat{\cal M}_{\alpha\beta}=\frac{\partial\sqrt{{\cal M}_{\alpha}}}{\partial
  k^{\mu}_{\alpha}}k_{\beta\mu}=-\frac{4\sqrt{{\cal M}}\Lambda^2}{(
\Lambda^2+k^2_{\alpha})^2} (k_{\alpha}\cdot k_{\beta}).
\end{equation}
%EQUATION<<<
For more details of evaluating Eq.~(\ref{eq:local}), one can
refer to Ref.~\cite{Nam:2007gf}.  

Since the pion EM form factor derived from the instanton vacuum is
only valid in the space-like region, we need further theoretical
consideration in order to connect the present calculation to the
change of the $\rho$-meson mass.  It is well known that the vector
meson dominance (VMD) describes the pion EM form factor quantitatively
in terms of $\rho$-meson exchange with the help of dispersion 
theory~\cite{Sakurai,Feynman:1973xc}.  However, in order to apply the
VMD, we use a modified expression for the pion EM form factor in
medium written as follows~\cite{Feynman:1973xc}: 
%EQUATION>>>
\begin{eqnarray}
\label{eq:EMFFVMD}
F_\pi^*(Q^2)\approx \frac{{\cal C}^*\,m^{*2}_{\rho}}
{m_{\rho}^{*2} + Q^2+i\Gamma_{\rho}^*m_{\rho}^{*}},
\end{eqnarray}
where $m_{\rho}^*$ and $\Gamma_{\rho}^*$ denote the mass and full decay
width of the $\rho$ meson in medium, respectively.  The parameter
${\cal C}^*$ encodes the change of the pion EM form factor in medium.  
Note that from now on the quantities with asterisk are those in
medium.  In fact, the prefactor $\mathcal{C}$ can be written as 
\begin{equation}
  \label{eq:ratio}
\mathcal{C} = \frac{f_{\rho\pi\pi}}{f_\rho},  
\end{equation}
where $f_{\rho\pi\pi}$ and $f_\rho$ denote the strong coupling
constant for $\rho\to\pi\pi$ and the photon-$\rho$ meson coupling
constant, respectively.  In free space, the universality relation
$f_{\rho\pi\pi}=f_\rho$ in the VMD allows us to put $\mathcal{C}^*$
equal to $1$ so that the pion EM form factor may be properly 
normalized to be $F_\pi(0)=1$ at $Q^2=0$.  Moreover, the decay width
$\Gamma_\rho$ plays only an important role in the vicinity of
$-Q^2=m_\rho^2$, so that we will not consider its change here, since
we are interested in the pion EM form factor in the space-like region.

The pion EM charge radius is defined as follows: 
\begin{eqnarray}
\label{eq:EMCR}
\langle r^2\rangle^* = -6\frac{\partial F_\pi^*(Q^2)}{\partial
  Q^2}\Bigg|_{Q^2=0}. 
\end{eqnarray}
Using Eqs.~(\ref{eq:EMFFVMD}) and (\ref{eq:EMCR}), we have the
following relation:  
%EQUATION>>>
\begin{eqnarray}
\label{eq:REL}
\frac{m^*_{\rho}}{m_{\rho}}=\left[\frac{{\cal C}^*\langle
    r^2\rangle}{\langle r^2\rangle}^*\right]^{1/2}.
\end{eqnarray}
%EQUATION<<<
Thus, the in-medium EM charge radius can be written as 
a function of $\mu_q$:
\begin{eqnarray}
\langle r^2(\mu_q)\rangle^*
=\langle r^2(0)\rangle
{\cal C}^*(\mu_q)\left[\frac{m_{\rho}}{m^*_{\rho}(\mu_q)}\right]^2
\approx0.45\,{\rm fm}^2\times{\cal C}^*(\mu_q)
\left[\frac{m_{\rho}}{m^*_{\rho}(\mu_q)}\right]^2,
\end{eqnarray}
%EQUATION<<<
Note that $\langle r^2(0)\rangle$ is replaced by the
experimental value $0.45\,{\rm fm}^2$~\cite{Yao:2006px}. Thus,
the $\rho$-meson mass for the nonzero $\mu_q$ can be easily obtained
as follows:
\begin{eqnarray}
\label{eq:mrho}
m^*_{\rho}(\mu_q)=m_{\rho}\left[\frac{0.45\,{\rm fm}^2\times{\cal C}^*(\mu_q)}
{\langle r^2(\mu_q)\rangle^*}\right]^{1/2}.
\end{eqnarray}
Since we can explicitly compute $\langle r^2(\mu_q)\rangle^*$ and
${\cal C}^*(\mu_q)$ in the N$\chi$QM, we can easily relate the change
of the pion charge radius to that of the $\rho$-meson mass.  

%--------------------------------------------------
\section{Results and discussion}
%--------------------------------------------------
We discuss now the results obtained from the present work.  We first
want to mention that there is no room to play with parameters in the
N$\chi$QM from the instanton vacuum, since all parameters, in
particular, $\mathcal{M}_0$ and $\Lambda$ are fixed already by the
saddle-point equation.  Thus, we will present almost all the results
as functions of the quark chemical potential $\mu_q$ in the folloinwg.   
Note that the parameterized form factor in Eq.~(\ref{eq:MFD}) is only
valid in lower $\mu_q$.  As $\mu_q$ is getting close to the value of
the phase transition ($\mu_q\approx 300$ MeV), we find numerical
instabilities due to a singular behavior of the parameterized quark form
factor.  However, as far as we stay in lower $\mu_q$, we are safe from
such instabilities.      

\begin{figure}[t]
\includegraphics[width=10cm]{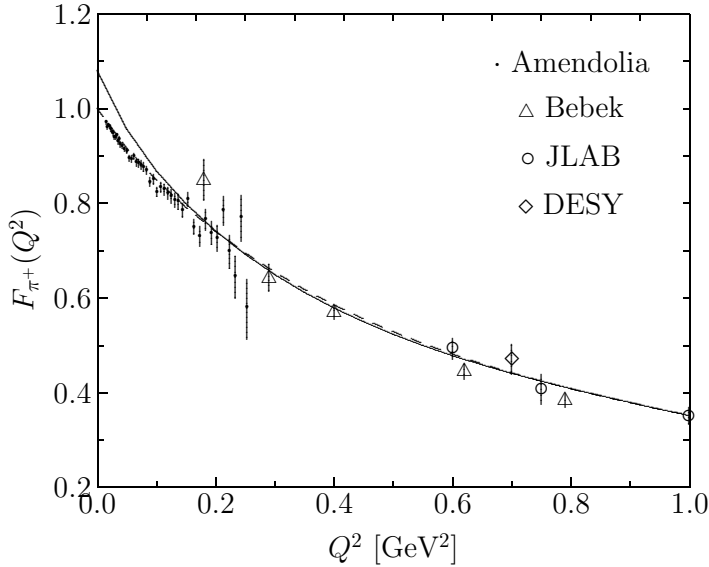}
\caption{The EM form factor of the pion, $F(Q^2)$ at finite density.  
The solid curve draws the pion EM form factor at $\mu_q=100$ MeV,
while the dashed curve represents that in free space.  The
experimental data in free space are taken from
Refs.~\cite{Bebek:1974iz,Bebek:1974ww,Amendolia:1986wj,Volmer:2000ek}.}       
\label{fig:2}
\end{figure}
In Fig.~\ref{fig:2}, we first show the results of the pion EM form
factor at $\mu_q=100$ MeV in comparison with that in free space.
While at higer $Q^2$ the pion EM form factor seems very similar to
that in medium, we observe that the pion EM form factor in the
vicinity of $Q^2=0$ becomes larger.  Moreover, the slope of the form
factor gets steeper near $Q^2=0$ with $\mu_q$ turned on.  This effect
will be clearly seen in calculation of the pion chage radius.   

\begin{figure}[t]
\includegraphics[width=10cm]{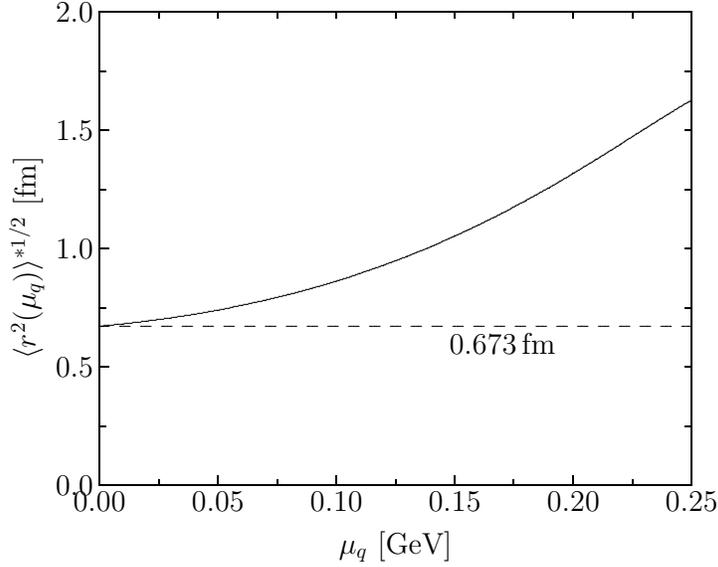}
\caption{The pion charge radius $\langle r^2\rangle^{*1/2}$ as a
function of the quark chemical potential $\mu_q$.  The solid curve
depicts the $\langle r^2\rangle^{*1/2}$, while the dashed line
represents that in free space, i.e. $\langle
r^2\rangle^{1/2}=0.673\,{\rm fm}$ (Experimental data: $\langle
r^2\rangle^{1/2}=0.672\pm0.008\,{\rm fm}$~\cite{Yao:2006px}).}
\label{fig:3}
\end{figure}
In Fig.~\ref{fig:3} we draw the numerical results for the $\langle
r^2\rangle^{*1/2}$ as a function of $\mu_q$.  It is shown that the
pion charge radius in medium increases as $\mu_q$ increases.  It
indicates, as mentioned already, that the pion EM form factor gets
steeper in the vicinity of $Q^2=0$ as $\mu_q$ increases.  One can
infer from this that the pion charge distribution gets smeared in
medium as the matter density gets denser.  In other words, the
interaction between the quarks inside the pion lose kinetic energies
in medium.  

\begin{figure}[t]
\includegraphics[width=10cm]{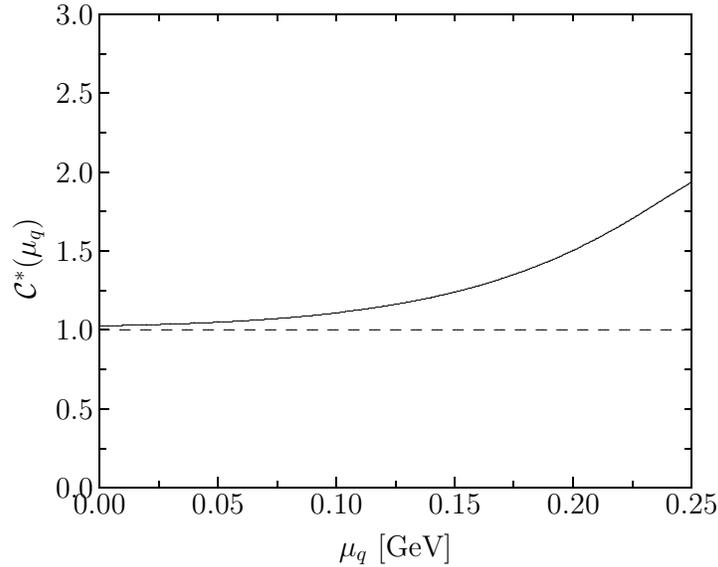}
\caption{The prefactor ${\cal C}^*(\mu_q)$ as a function of
$\mu_q$.  The solid curve represents the present result, while the
dashed line denotes its value in free space, i.e. $\mathcal{C}=1$.}  
\label{fig:4}
\end{figure}
Figure~\ref{fig:4} depicts the prefactor ${\cal
  C}^*(\mu_q)$ evaluated at $Q^2=0$ in the N$\chi$QM.  Being similar 
to the pion charge radius, the ${\cal C}^*(\mu_q)$ increases as
$\mu_q$ increases.  In order to understand the physical meaning of the
${\cal C}^*$, let us consider Eq.~(\ref{eq:ratio}) in which the ${\cal
  C}^*$ is given as a ratio $f_{\rho\pi\pi}^*$ and $f_\rho^*$.
It indicates from the result shown in Fig.~\ref{fig:4} that
$f_{\rho\pi\pi}^*$ and $f_{\rho}^*$ in medium are modified in
different ways, though we cannot tell how they are changed
explicitly. 

\begin{figure}[t]
\includegraphics[width=10cm]{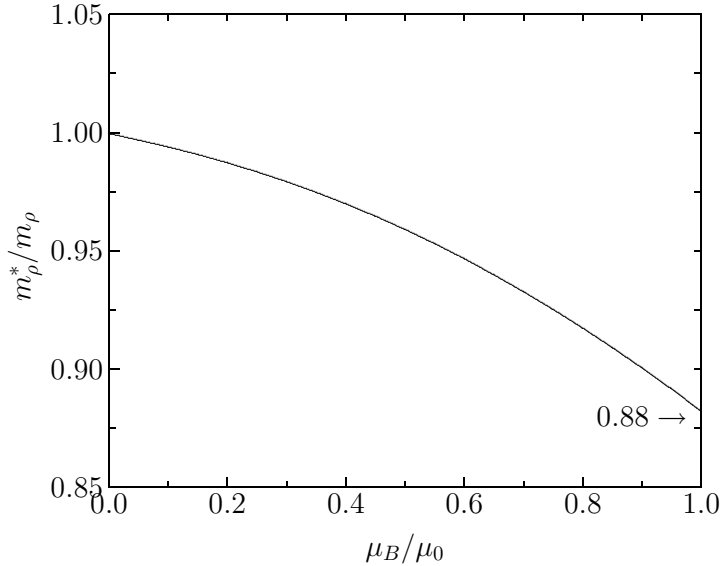}
\caption{The ratio of the $\rho$-meson mass $m_\rho^*/m_\rho$ as a
function of $\mu_B/\mu_0$.  The baryon chemical potential $\mu_B$ is
defined as $\mu_B=3\mu_q$.  The $\mu_0$ denotes the chemical potential
corresponding to normal nuclear matter density
$\rho_0\approx0.17\,\mathrm{fm}^{-3}$.}  
\label{fig:5}
\end{figure}
Finally, we are in a position to discuss the mass shift of the $\rho$
meson.  Inserting the results of $\langle r^2(\mu_q)\rangle^*$ and
${\cal C}^*(\mu_q)$ as already shown in Figs.~\ref{fig:3} and
\ref{fig:4}, respectively, into Eq.~(\ref{eq:mrho}), we arrive at the
final result of the $\rho$-meson mass shift in medium as shown in 
Fig.~\ref{fig:5}.  Obviously, the $\rho$-meson mass decreases as the
$\mu_q$ increases.  This is due to the fact that the result of the
pion charge radius squared, $\langle r^2(\mu_q)\rangle^*$, grows
faster than that of the ${\cal C}^*(\mu_q)$.  As a result, the
$\rho$-meson mass, which is proportional to a ratio of $\sqrt{{\cal
C}^*/\langle r^2\rangle^*}$ as shown in Eq.~(\ref{eq:mrho}), falls off
approximately linearly as $\mu_q$ increases.    

Since the $\rho$-meson mass decreases approximately linearly, we are
able to employ the following linear parameterization for the mass shift
of the $\rho$ meson as in Ref.~\cite{Hatsuda:1991ez}: 
%EQUATION>>>
\begin{eqnarray}
\label{eq:para}
\frac{m^*_{\rho}}{m_{\rho}}\approx 1-\alpha\,\frac{\mu_B}{\mu_0},
\end{eqnarray}
%EQUATION<<<
where the $\mu_B$ denotes the baryon number chemical potential being
equal to $3\mu_q$.  The $\mu_0$ denotes the chemical potential
corresponding to normal nuclear matter density, $\rho_0\approx0.17\,{\rm
fm}^{-3}$, and is chosen to be about $300$ MeV.  We note that $\mu_0$
was estimated to be $308$ MeV, the nucleon mass and its binding energy
being taken into account~\cite{Buballa:2003qv}.  
$\alpha$ is a coefficient for showing the mass dropping of the $\rho$
meson.  In fact, the  mass ratio of Eq.~(\ref{eq:para}) is usually 
parameterized as a function of a ratio of the nuclear matter densities 
$\rho_B/\rho_0$~\cite{Brown:1991kk,Hatsuda:1991ez}: 
$m^*_{\rho}/m_{\rho} \approx 1-\alpha\,\rho_B/\rho_0$.  However,
in the lower $\mu_q$ region ($\mu_q\lesssim200$ MeV), the baryon
number density turns out to be zero due to the following relation
$\rho_B\propto[\mu_q^2-M^2_q]^{3/2}$ in which $M_q$ is the constituent
quark mass, $M_q=300\sim400$ MeV~\cite{Carter:1998ji}.  Hence, we only
show in the present work the $\rho$-meson mass shift as a function 
of $\mu_B/\mu_0$.  The value of $\alpha$ is then determined to be
$\alpha\approx 0.12$, i.e. $m^*_{\rho}/m_{\rho}\approx 0.88$.  The
present result is slightly smaller than
Refs.~\cite{Brown:1991kk,Hatsuda:1991ez}: In 
Ref.~\cite{Hatsuda:1991ez}, $\alpha$ was estimated as $0.15\sim0.18$,
which is similar to that based on phenomenological method in
Ref.~\cite{Brown:1991kk}.  We want to mention also that,
in contrast, the $\rho$-meson mass drops suddenly as the $\mu_q$
increases in Ref.~\cite{Muroya:2002ry} using the lattice simulation in
color SU(2) symmetry.  The present result $\alpha\approx 0.12$ is 
also consistent with a recent measurement at KEK~\cite{Naruki:2005kd}.
%--------------------------------------------------
\section{Summary and conclusion}
%--------------------------------------------------
We have investigated the pion electromagnetic form factor and pion
charge radius at finite density within the framework of the nonlocal
chiral quark model from the instanton vacuum.  The pion
electromagnetic form factor is getting steeper near $Q^2=0$ as $\mu_q$
increases.  As a result, the pion charge radius grows as $\mu_q$
increases.  With the help of the vector meson dominance, we were able
to express the $\rho$-meson mass shift in terms of the prefactor,
which is written as a ratio of the $\rho$-$\pi$-$\pi$ strong coupling
constant and $\rho$-photon constant, and pion charge radius.  

In the present work, we have observed that the pion charge radius
increases with respect to the quark chemical potential which implies that
the pion charge distribution is smeared in medium as the density gets
denser and the interaction between quarks gets lessened.  The
prefactor $\mathcal{C}$ also is getting stronger as $\mu_q$
increases.  With these two quantities, we have shown that 
the $\rho$-meson mass falls off as $\mu_q$ increases, which is similar
to that suggested by Brown-Rho scaling law.  When we made a linear 
parameterization for the mass shift ($m^*_{\rho}/m_{\rho} \approx
1-\alpha\,\mu_B/\mu_0$), the coefficient $\alpha$ turns out to be
about $0.12$, which is compatible to those of other models and
phenomenological estimations ($0.15\sim0.18$).  Moreover, the present
result is consistent with a recent
measurement of the in-medium $\rho$-meson mass
shift~\cite{Naruki:2005kd}.   

We finally stress that the mass shift of the $\rho$-meson can be also
tested by measuring the pion EM charge radius in medium.  Thus, we
anticipate experiments focussing on the pion EM form factor in  
medium ($T\approx0$ and $\mu_q\ne0$) to test the present theoretical
assertion.
%-------------------------------------------------
\section*{Acknowledgments}
%-------------------------------------------------
The authors would like to thank T.~Kunihiro, C.~H.~Lee, S.~H. Lee, M.~Harada,  
M.~M.~Musakhanov, and I.~K.~Yoo for stimulating and fruitful
discussions.  The work of S.i.N. is partially supported by the grant
for Scientific Research (Priority Area No. 17070002) from the Ministry
of Education, Culture, Science and Technology, Japan. The work of
H.Ch.K. is supported by the Korea Research Foundation Grant funded by
the Korean Government(MOEHRD) (KRF-2006-312-C00507).  The numerical
calculations were carried out on YISUN at YITP in Kyoto University and
on MIHO at RCNP in Osaka University. 
%-------------------------------------------------- 

\end{document}